\documentclass[%
preprint,
superscriptaddress,
 amsmath,amssymb,
 aps,
prl,
]{revtex4-2}

\usepackage{graphicx}% Include figure files
\usepackage{dcolumn}% Align table columns on decimal point
\usepackage{bm}% bold math
\usepackage{hyperref}% add hypertext capabilities
\usepackage{color}
\usepackage{lipsum}
\usepackage{multirow}
\usepackage{array,booktabs}
\usepackage{soul}
\newcolumntype{M}[1]{>{\centering\arraybackslash}m{#1}}

\newcommand{\bfrm}{\boldsymbol{\mathbf{r}}^N}
\newcommand{\mr}{\mathbf r}
\newcommand{\mx}{\mathbf x}
\newcommand{\mF}{\mathbf F}

\begin{document}

% \preprint{APS/123-QED}
\title{Modulating internal transition kinetics in responsive macromolecules by collective crowding}
% \title{dada}

\author{Upayan Baul}
\affiliation{Applied Theoretical Physics - Computational Physics, Physikalisches Institut, Albert-Ludwigs-Universit\"at Freiburg, D-79104 Freiburg, Germany}
\author{Nils G\"{o}th}
\affiliation{Applied Theoretical Physics - Computational Physics, Physikalisches Institut, Albert-Ludwigs-Universit\"at Freiburg, D-79104 Freiburg, Germany}
\author{Michael Bley}
\affiliation{Applied Theoretical Physics - Computational Physics, Physikalisches Institut, Albert-Ludwigs-Universit\"at Freiburg, D-79104 Freiburg, Germany}
\author{Joachim Dzubiella}
\affiliation{Applied Theoretical Physics - Computational Physics, Physikalisches Institut, Albert-Ludwigs-Universit\"at Freiburg, D-79104 Freiburg, Germany}
\affiliation{Cluster of Excellence livMatS @ FIT - Freiburg Center for Interactive Materials and Bioinspired Technologies, Albert-Ludwigs-Universit\"at Freiburg, D-79110 Freiburg, Germany}

\date{\today}

\begin{abstract}
Packing and crowding are used in biology as mechanisms to (self-)regulate internal molecular or cellular processes based on collective signalling. Here, we study how the transition kinetics of an internal 'switch' of responsive macromolecules is modified collectively by their spatial packing. We employ Brownian dynamics simulations of a model of 'responsive colloids' (RCs), in which an explicit internal degree of freedom -- here, the particle size -- moving in a bimodal energy landscape responds self-consistently to the density fluctuations of the crowded environment. We demonstrate that populations and transition times for the two-state switching kinetics can be tuned over one order of magnitude by 'self-crowding'. An exponential scaling law derived from a combination of Kramers' and liquid state perturbation theory is in very good agreement with the simulations. 
% We study how the the transition kinetics of internal degrees of freedom of responsive macromolecules are modified by their spatial packing, using particle-resolved Brownian dynamics computer simulations and a combination of Kramers' and liquid state perturbation theory. We employ a recently developed model of 'responsive colloids' (RCs) in which an explicit internal degree of freedom responds self-consistently to the collective density fluctuations in the crowded environment. In our case study, we focus on the two-state switching kinetics of the size (small and large) of the macromolecule. We find that the transition times between the two sizes are modified by self-crowding by one order of magnitude. Our theory is in very good agreement with the simulated kinetics and provides general scaling laws which should be easily testable in various experiments. 
% 
\end{abstract}
%\keywords{Suggested keywords}%Use showkeys class option if keyword display desired
\maketitle
%\cmt{All older text retained. Please see commented text in the tex file.}\\
%\nils{Nils' color}
\clearpage
Over the last two decades the phenomenon of macromolecular crowding has attracted great attention because of its large effects on biomolecular mobility, reactions, structure and dynamics~\cite{Ellis2001,minton, Rosalind, MittalPrl2013, ThirumalaiPrl2015}. In particular, crowding leads to variations of (two-state) folding energy landscapes and kinetics, typically leading to more compact molecular states and speeding up folding times~\cite{ThirumalaiPnas2005, Gierasch, Dupuis, Metzler:SoftMatter}.  In fact, it was proposed that cells can regulate and maintain a relatively constant level of packing of macromolecules to tune molecular processes, which was termed 'homeocrowding' as an acronym of macromolecular crowding homeostasis~\cite{homeocrowding}. 
% Hence, biology has the ability to tune the internal function of a (bio)macromolecule by a collective density control. 

Similar mechanisms play decisive roles also on larger scales, for example, in bacterial 'quorum sensing', which is the self-regulation of internal gene expression rates in response to fluctuations in overall cell-population density~\cite{quorum}. These fascinating collective effects are inspiring for the design of 'intelligent' synthetic materials, where soft and responsive particles, e.g., polymer-based hydrogels, potentially mimic these effects for discovering new functionalities, such as advanced time responses, feedback and adaptivity~\cite{Heuser,Andreas}. In particular, the colloid/polymer duality of soft microgels~\cite{NievesAnnRevPhysChem2012,Keidel} and their stimuli-triggered conformational switching~\cite{Stuart2010} open new avenues for creating programmable, self-regulating materials. 

Despite the progress in this field, our understanding is still based on relatively simple soft matter concepts such as excluded volume, depletion interactions, or scaled particle theory~\cite{BestBiophysJour2010,homeocrowding}. In other words, in conventional theoretical studies of macromolecular crowding the crowders are hard and inert, while in reality they are soft and responsive. Like in bacteria, control by crowding is collective and self-consistent, affecting the crowder and the crowded simultaneously~\cite{quorum}. Physical models that self-consistently address crowding effects and their consequence on internal processes, such as internal transitions or reactions, are still absent. A promising approach in this regard is the explicit consideration of internal degrees of freedom (DOFs) which can respond to environmental changes~\cite{matthias,LinPRE,briels,Lerner19_PRE,Singh19_JSM,Loewen21_Lang}, including local density fluctuations~\cite{Upi}.

% Physical models that address self-consistently crowding effects (including the effect of local density fluctuations) and their consequence on internal processes, such as internal transitions or reactions, are still absent. 

%%%% UB : The following line is removed because it has been said multiple times in the above. Seems redundant at this point.
% Here, we introduce such a self-consistent methodology to study the effects of collective crowding on internal processes in responsive macromolecules. 
In this letter, we study collective crowding effects for a single macromolecular species ('self-crowding') employing a model of Responsive Colloids (RCs)~\cite{LinPRE,Upi}. The latter explicitly resolves an internal 'property' DOF that responds self-consistently to changes in the local packing. As a case study, we focus on the two-state switching kinetics of the size of the macromolecule, modeling, for example, conformational transitions in globular protein folding~\cite{ThirumalaiPnas2005,Gierasch,Dupuis}, switching of intrinsically disordered proteins (IDPs) \cite{bowen} or synthetic supramolecular polymers~\cite{helical_switching}, or bimodal hydrogel volume transitions~\cite{Heuser,oscillating,DNA_hydrogel, breathing}. We find that populations and the transition times between the two sizes (small and large) modified by packing density spread over one order of magnitude. Importantly, we develop a scaling theory based on a combination of Kramers' escape~\cite{BorkovecRevModPhys1990} and liquid state perturbation theory~\cite{HansenMcDonaldbook} extended to RCs. The theory is in very good agreement with the calculated kinetics and provides useful, experimentally testable scaling laws for a general class of Hamiltonians. 

% 
% We consider only a single species ('self-crowding'), but the method can easily be extended to multiple macromolecules in the suspension. We employ our recently introduced model of 'responsive colloids' (RCs)~\cite{LinPRE,Upi} which explicitly resolves internal 'property' degrees of freedom that respond self-consistently to the collective density fluctuations in the environment. As our case study, we focus on the internal two-state switching kinetics of the size of the macromolecule, modeling, for example, two-state protein folding~\cite{ThirumalaiPnas2005} or bimodal hydrogel volume transitions (references in \cite{moncho}) and systematically examine its dependence on the particle packing. We find that the calculated transition times between the two sizes (small and large) are modified by crowding by one order of magnitude. Importantly, we develop a theory based on a combination of Kramers' escape~\cite{BorkovecRevModPhys1990} and liquid state perturbation theory~\cite{HansenMcDonaldbook} extended to RCs. Our theory, which is generally applicable to arbitrary RC Hamiltonians, is in very good agreement with the calculated kinetics and provides useful scaling laws. 

\begin{figure}[t!]
 \centering
 \includegraphics[width=450pt]{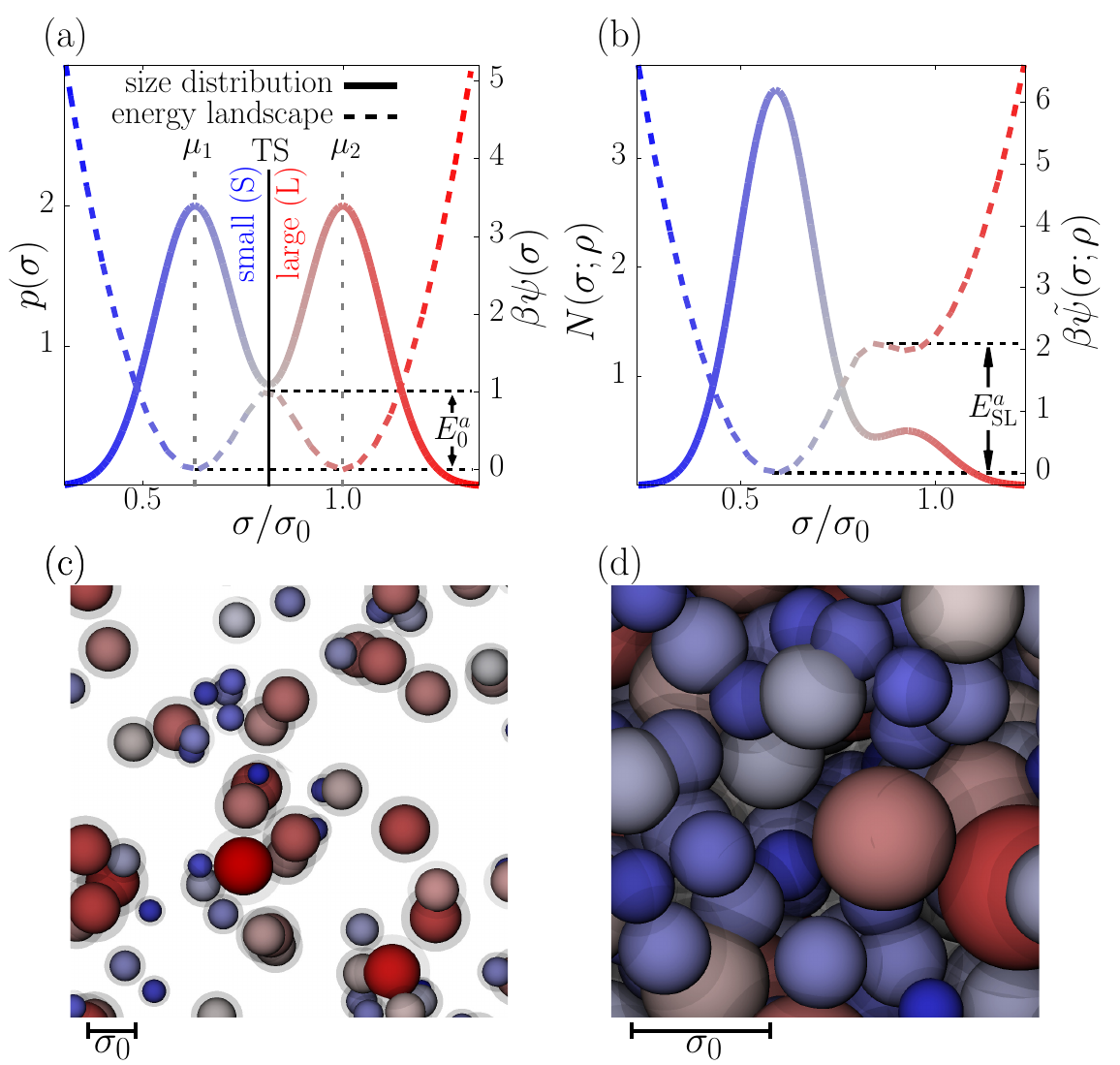}
% \resizebox{!}{0.65\textwidth}{\input{Figures/texFig_1.tex}}
 \caption{{\footnotesize 
 (a) Parent distribution $p(\sigma)$ of size $\sigma$  (solid) and corresponding energy landscape $\psi(\sigma)$ (dashed) at dilute (single-particle) conditions. Locations of energy minima ($\mu_1, \mu_2$) and the corresponding transition state (TS) are shown, separating small (S, blue) and large (L, red) states. (b) Example emergent distribution $N(\sigma;\rho)$ (solid) and corresponding energy landscape $\tilde \psi(\sigma;\rho)$ (dashed). The crowding changes the activation energy from $E_0^a$ in (a) to $E_{\rm SL}^a$ in (b).  (c, d) Representative snapshots of system configurations under dilute and crowded conditions, respectively. The color gradient from blue (small sizes) to red (large sizes) visualizes the dynamical sizes according to the distributions in the top panels (note the scale bars).}}

 \label{fgr:fig1}
\end{figure}

We consider $N$ interacting particles in a volume $V$ with number density $\rho=N/V$ in absence of any external fields. The coarse-grained RC Hamiltonian is defined as~\cite{LinPRE}
\begin{equation}
H ( \bfrm, \sigma^N ) = \sum_i^N \psi(\sigma_i) + \frac{1}{2} \sum_{i\neq j}^N \phi(\mr_i,\mr_j; \sigma_i,\sigma_j),
\label{eqn:FreeEnergy}
\end{equation}
and describes the energy of the microstate of the configuration \{$\mr_i,\sigma_i$\}$=:\{\mx_i\}$, where the $\mr_i$ and $\sigma_i$ are positions and sizes of particle $i=1..N$, respectively. The first term on the right-hand-side carries the one-body contribution to the energy associated with changes in the size $\sigma$. Thus, it represents the energy landscape
$ \beta \psi (\sigma)= -\ln\,p(\sigma)$,
where $p(\sigma)$ is the corresponding probability distribution, and $\beta = 1/k_BT$ the inverse thermal energy. We refer to $p(\sigma)$ as the \textit{parent} distribution since it describes the intrinsic internal polydispersity of an isolated particle. To model two-state kinetics, we choose a bimodal form of the explicit, double-Gaussian form 
\begin{eqnarray}
 p(\sigma) = \frac{1}{\sqrt{2\pi}} \biggl[ \frac{c}{\delta_1} e^{\left( -\frac{1}{2} \left( \frac{\sigma - \mu_1}{\delta_1} \right)^2 \right)}  
 + \frac{1 - c}{\delta_2} e^{\left( -\frac{1}{2} \left( \frac{\sigma - \mu_2}{\delta_2} \right)^2 \right)} \biggr] .
 \label{eqn:parent}
\end{eqnarray}
This and its corresponding energy landscape $\psi(\sigma)$ are shown in Fig.~\ref{fgr:fig1}(a), highlighting the energy minima $\mu_1$, $\mu_2$, and the local maximum, defining the transition state (TS). In going from continuous distributions to the two-state (small (S) versus large (L)) picture, we consistently use the TS as the state boundary. In our study, $c = 0.5$, $\mu_1 = 0.63\sigma_0$, $\mu_2 = 1.0\sigma_0$, and $\delta_1=\delta_2\equiv\delta = 0.1\sigma_0$, with unit length $\sigma_0$ set to 1.0. The choice leads to an unperturbed activation energy barrier $E_0^a \simeq 1 k_BT$ (Fig.~\ref{fgr:fig1}(a)).

The second term in Eq.~(\ref{eqn:FreeEnergy}) represents the two-body, effective pair potential, $\phi(\mr_i,\mr_j; \sigma_i,\sigma_j)$, which is explicitly dependent on both the positions and the property of the interacting RCs. Here, we choose the Hertzian potential
\begin{equation}
 \phi(r; \sigma_i,\sigma_j) = \epsilon \left( 1 - \frac{r}{\sigma_{ij}} \right)^{5/2} \Theta \left( 1 - \frac{r}{\sigma_{ij}} \right),
 \label{eq:pp}
\end{equation}
where $\Theta$ is the Heaviside step function, $\epsilon$ dictates the magnitude of the interaction, $r=|\mr_j - \mr_i|$ is the interparticle distance, and $\sigma_{ij}=(\sigma_i+\sigma_j)/2$. The Hertzian potential reflects the soft, purely repulsive interactions between spherical elastic colloids with diameter $\sigma_i$, widely used to describe swollen microgels~\cite{ZaccMacromol2019,DentonSM2016,SchurtNatComm2018,SchurtJCP2014}. For our RC system we choose $\beta \epsilon=500$, which is in the typical range for modeling real experimental systems~\cite{SchurtNatComm2018}. 
% \cmt{Specify $\sigma_{ij}$ and more, if needed.}. 

% The equilibrium distribution of sizes of the interacting system is defined as $N(\sigma;\rho) = \langle \sum_{i=1}^N \delta(\sigma-\sigma_i)\rangle$, $\langle..\rangle$ being the canonical ensemble average. The distribution $N(\sigma;\rho)$ is normalized to unity by dividing by the total number of particles $N$. Importantly, $N(\sigma;\rho)$ represents the \textit{emergent} property distribution accounting for interactions among RCs and is a central property of our study, because 
The equilibrium probability distribution of sizes in the dispersion is defined as $N(\sigma;\rho) = \frac{1}{N} \langle \sum_{i=1}^N \delta(\sigma-\sigma_i)\rangle$ by the canonical ensemble average. It represents the \textit{emergent} property distribution accounting for the effects of interactions among RCs. It is a central in our study, because 
\begin{equation}
\beta \tilde \psi(\sigma;\rho) = -\ln N(\sigma;\rho)
\label{eqn:PsiTildeSigma}
\end{equation}
defines the crowding-modified energy landscape, which tunes the internal two-state switching. Fig.~\ref{fgr:fig1}(b) exemplifies the modified energy barrier $E_{\mathrm{SL}}^a$ for S to L transitions. Since we have a Hamilton system with explicit response of the sizes, the modification of the energy landscape is collective and self-consistent.

We assume both the  translation and the size dynamics to be overdamped and  propagated by Brownian dynamics (BD)~\cite{BerendsenBook}, as 
% \begin{equation}
%  \mx_i(t+\Delta t) = \mx_i(t) + \frac{\bf D}{k_BT} \mF_i(t) \Delta t + \bm{\xi},
%  \label{eqn:BDtrans}
% \end{equation}
% where $\Delta t$ is the simulation timestep and ${\bf D} = (D_i, D_i, D_i, D_\sigma)$ represents the diffusion coefficients, which are all the same for the three translation directions. 
\begin{equation}
 \mx_i(t+\Delta t) = \mx_i(t) + \frac{D_{l,i}}{k_BT} \mF_i(t) \Delta t + \bm{\xi}_i,
 \label{eqn:BDevol}
\end{equation}
where $\Delta t$ is the simulation timestep.  $\mF_i = -\nabla_i H$ is the generalized force experienced by $i$-th particle from pairwise interactions given Hamiltonian $H$, Eq.~(\ref{eqn:FreeEnergy}), and $\nabla= (\partial/\partial x, \partial/\partial y, \partial/\partial z, \partial/\partial \sigma)$ 
% $\nabla= \sum_{\alpha=x,y,z} \partial/\partial \alpha_i + \partial/\partial \sigma$ 
is a 4-gradient. 
The term $\bm{\xi}_i$ represents stochastic forces stemming from the solvent fluctuations. For the evolution along a coordinate $l=x,y,z,\sigma$, the corresponding component $\xi_l$ is drawn from a Gaussian distribution with mean $\langle \xi_l \rangle = 0$ and variance $\langle \xi_l^2 \rangle = 2 D_l \Delta t$, which is strictly $\delta$-correlated in time and obeys the standard fluctuation-dissipation theorem~\cite{AllenTildesley}. For translation ($l=x,y,z$), the diffusivity $D_{l,i}$ in Eq.~(\ref{eqn:BDevol}) is defined in units $\sigma_0^2/\tau_{\rm BD}$, where $\tau_{\rm BD}$ sets our unit (Brownian) time scale.  The diffusion is related to the friction coefficient $\zeta_{l,i}$ through $D_{l,i} = k_BT / \zeta_{l,i}$. We assume the latter to follow the Stokes-Einstein relation $\zeta_{l,i} \propto \sigma_i$, as recently verified for conformationally fluctuating proteins~\cite{PhysRevLett.126.128101}. 
% \cmt{To set the time scale of size fluctuations, we assume $D_\sigma$...? Introduce $\alpha$ here. Mention relevance for Kramers already. Table of systems and parameters? Summarize briefly some simulations details, e.g., $N$, box size, run length, etc.}
Setting the time scale of size ($\sigma$) fluctuations, we define $D_\sigma = \alpha\, \sigma_0^2/\tau_{\rm BD}$ and use the parameter $\alpha$ (same for all particles) to tune the intrinsic time scale of internal switching. 

We simulate fully periodic systems in the canonical ($NVT$) ensemble comprising $N=512$ RCs for $\alpha = (0.001, 0.01, 0.1, 1.0, 10.0)$ and particle densities in the range from the low density limit (LDL), $\rho \sigma_0^3 \to 0$, up to $\rho \sigma_0^3=1.91$. The LDL is approximated by simulating an isolated particle. More simulation details are given elsewhere~\cite{SI}. Simulation snapshots with particle size distributions resolved by color are illustrated in Figs.~\ref{fgr:fig1}(c) and (d). 

% Emergent size distributions $N(\sigma;\rho)$ for various simulated densities $\rho$ are presented in FIG.~\ref{fgr:fig2}~(a). For larger packing, the particles are squeezed and the bimodal distribution shifts to a more stable smaller size. Due to the - on average - smaller size of the particles, the first peak of the radial distribution functions, shown in FIG.~2(b), moves to smaller particle distances. The first peak is also growing as expected for larger packing, although the decreasing particle size is counterbalancing this effect. This can be rationalized by the effective packing fraction.. \cmt{Should we report effective packing fractions somewhere, e.g. , based on the mean size? They should increase less rapidly than density. Could be another plot in a 2x2 FIG. 2, including $\tilde \Psi$. Or a table.}
% \cmt{current 2b could alternatively be an inset to 2a. We should move 3a into FIG.2}. The crowding-modified landscape $\beta \tilde \Psi(\sigma;\rho) = -\ln N(\sigma;\rho)$ is shown in FIG.~2.... Small sizes are substantially stabilized over the larger states with increasing density. Importantly, however, the transition state is also drastically modified which is the key to characterize the changes in the transition kinetics. 

\begin{figure*}[tp!]
 \centering
 \includegraphics[width=450pt]{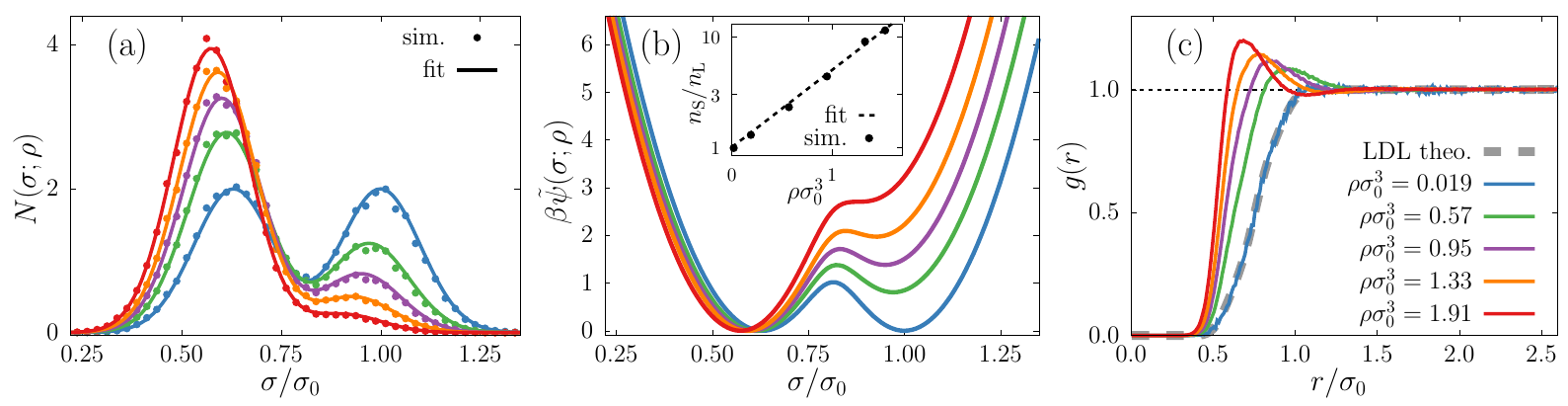}
% \resizebox{!}{0.25\textwidth}{\input{Figures/texFig_2_fit.tex}}
 \caption{{\footnotesize (a) Emergent size distributions $N(\sigma;\rho)$, (b) energy landscapes $\tilde \psi(\sigma;\rho)$, and (c) RDF $g(r)$  for varying number densities $\rho$. The distributions in (a) can be well fitted to bimodal distributions of form of Eq.~(2)~\cite{SI}, which we take to compute $\tilde \psi(\sigma;\rho)$ in (b) using Eq.~(\ref{eqn:PsiTildeSigma}). The inset in (b) shows the population ratio between small (S) and large (L) states versus $\rho$, calculated from integrations over the distribution, $n_S/n_L = \int_{-\infty}^{\rm TS}N(\sigma){\rm d}\sigma / \int_{\rm TS}^\infty N(\sigma){\rm d}\sigma$. The dashed gray line in (c) shows the theoretical estimate in the LDL given by $g(r) \approx \int \mathrm{d}\sigma' \int \mathrm{d}\sigma p(\sigma') p(\sigma) \exp{\left[-\beta \phi(r; \sigma, \sigma')\right]}$.}}
%  $g(r) \approx \int_{-\infty}^{\infty} \mathrm{d}\sigma' \int_{-\infty}^{\infty} \mathrm{d}\sigma \exp{\left[-\beta \phi(r; \sigma, \sigma')\right]}$.
 \label{fgr:fig2}
\end{figure*}

Resulting emergent size distributions $N(\sigma;\rho)$ are presented in Fig.~\ref{fgr:fig2}(a). At dense packing, the particles are compressed and the bimodal distribution shifts to stabilize smaller sizes. For all densities, the $N(\sigma;\rho)$ distributions also fit well to bimodal Gaussians (Eq.~(\ref{eqn:parent}))~\cite{SI}. Fig.~\ref{fgr:fig2}(b) shows the crowding-modified landscapes $\beta \tilde \psi(\sigma;\rho) = -\ln N(\sigma;\rho)$ computed using the fitted $N(\sigma;\rho)$ distributions. Small sizes are substantially stabilized over the larger states with increasing density, with small-to-large population ratios exceeding a factor of 10, see the inset of Fig.~\ref{fgr:fig2}(b).  This is in agreement with earlier computational studies which reported crowding induced coil-to-globule transitions of two-state biopolymer systems consistent with experiments~\cite{ThirumalaiPrl2015}.  Importantly, the transition barriers are drastically modified by packing, which is the key to characterize the changes in the transition kinetics.

For further analysis, we introduce the two-body density-size distribution function~\cite{LinPRE},
%\begin{equation}
% \rho^{(2)}(\mathbf r,\mathbf r', \sigma, \sigma') = \left\langle \sum_{i=1}^N \sum_{j=1}^N \delta(\mathbf r - \mathbf r_i)\delta(\mathbf r' - \mathbf r_j)\delta(\sigma-\sigma_i)\delta(\sigma'-\sigma_j)\right\rangle
% \label{eqn:PropDen2Body}
%\end{equation}
$g(\mx,\mx')\equiv g(r; \sigma, \sigma')$. By integrating out $\sigma$ and $\sigma'$~\cite{LinPRE}, we recover the usual radial distribution function (RDF) $g(r)$, which represents the average over the polydisperse distributions. Due to the - on average - smaller size of the particles, the first peak of the RDF, shown in Fig.~\ref{fgr:fig2}(c), shifts to smaller interparticle distances with increasing $\rho$. The first peak grows for higher density as expected from enhanced packing correlations, although the decreasing particle size counterbalances this effect. This can be rationalized by the effective packing fraction, defined as $\eta = \pi\rho\langle\sigma^3\rangle/6$~\cite{Upi}, which approaches sublinear behavior with increasing $\rho$~\cite{SI}. Such a weak scaling with density may indeed facilitate homeocrowding as hypothesized for biological cells~\cite{homeocrowding}.
% \cmt{Should we report effective packing fractions somewhere, e.g. , based on the mean size? They should increase less rapidly than density. Could be another plot in a 2x2 FIG. 2, including $\tilde \Psi$. Or a table.}
% \cmt{Should we merge conceptual figure 3(b) with a plot in FIG.1, e.g. current 1c? I think one can do that.} 

\begin{figure*}[ht!]
 \centering
 \includegraphics[width=450pt]{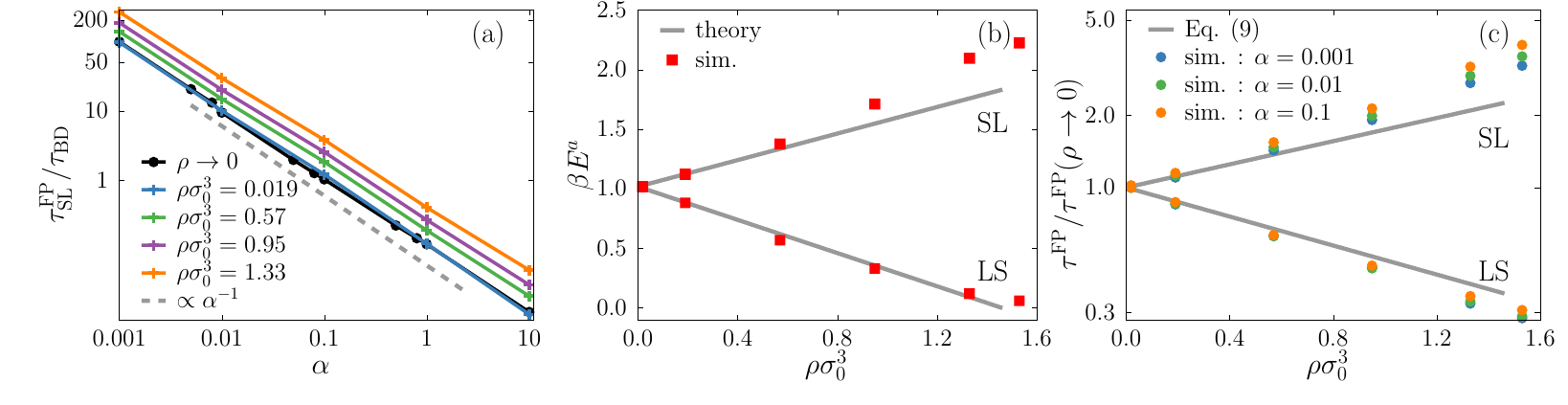}
% \resizebox{!}{0.25\textwidth}{\input{Figures/texFig_3.tex}}
 \caption{{\footnotesize (a) Simulation MFPTs of small to large (SL) size switching (in units of the Brownian time scale $\tau_{\rm BD}$) versus $\alpha$ at various densities shown using log-log scale. The dashed gray line denotes the $\alpha^{-1}$ Kramers' scaling behavior, Eq.~(\ref{eqn:KramersRule}). (b) Comparison of theory (lines) and simulations (sim., red squares) of the activation energy $E^{a}$ versus density $\rho$. (c) Comparison of theory and simulations of the MFPT (scaled by the LDL) as a function of $\rho$, using a log-lin scale.}}
 \label{fgr:fig3}
\end{figure*}

From our simulations we directly access the mean first passage times (MFPTs) $\tau^{\rm FP}$ for {\it small to large} (SL) or {\it large to small} (LS) transitions from time series analysis~\cite{mfpt,SI}. For our overdamped systems, Kramers' escape theory applies~\cite{BorkovecRevModPhys1990}, which essentially reads
% \begin{equation}
% \tau = \frac{(\mu_2-\mu_1)^2}{\alpha}\exp(-\beta E^a), 
% \end{equation}
\begin{equation}
\tau^{\rm FP} (\alpha,\rho) \sim \alpha^{-1} \exp(-\beta E^a(\rho)),
\label{eqn:KramersRule}
\end{equation}
where $E^a(\rho)$ is the energy barrier obtained from $\tilde \psi(\sigma;\rho)$, cf. Fig.~2(b). To check the prefactor scaling with $\alpha$, we plot first the SL-MFPT, $\tau^{\rm FP}_{\rm SL}$, as a function of $\alpha$ in Fig.~\ref{fgr:fig3}(a). Indeed, we observe a clear $\alpha^{-1}$ dependence for all densities, that is, the intrinsic time scale of $\sigma$ simply scales the MFPT linearly. 

The density-dependence of the MFPTs, though, is non-trivial and dictated by self-crowding induced changes to $\tilde \psi(\sigma;\rho)$, and consequently the changes in the activation energies $E^a(\rho)$  (see Figs.~\ref{fgr:fig1}(a,b) and~\ref{fgr:fig2}(b)). In particular, the SL transition slows because of an increasing energy barrier with density, while the LS transition accelerates. The calculated energy barriers are presented in Fig.~\ref{fgr:fig3}(b). They depend apparently linearly on the density, changing by up to a factor 2-3, due to the compression of the energy landscapes.  This change of activation energy results in a drastic, almost one-order-of magnitude spread of all MFPTs versus density, as presented in Fig.~\ref{fgr:fig3}(c) as a log-lin plot. We observe that the MFPTs depend exponentially on the crowding density.   [We note here that we find, as expected, the exponential relationship between times $\tau^{\rm FP}$ and energy $E^a$ holds across all $\rho$~\cite{SI}, as predicted by Kramers', Eq.~(\ref{eqn:KramersRule}).] At a critical density of about $\rho\sigma_0^3 \simeq 1.7$ the transition vanishes in our systems because the L-state becomes unstable, as indicated by the saddle point in the distribution. Here, $\tau^{\rm FP}_{\rm LS}$ accordingly approaches zero, which can be interpreted as a completely stabilized  (folded or globule) S-state. 

The LS-transition can be identified with a folding transition from coil to globule states~\cite{Gierasch,Dupuis}; qualitatively consistent with our results, it was demonstrated experimentally and through simulations  that crowding enhances biomolecular folding rates by a multiple, while unfolding rates decrease~\cite{ThirumalaiPnas2005, Dupuis}. Similarly, large crowders speed up two-state polymer looping kinetics in DNA models~\cite{Metzler:SoftMatter}. Note, however, that the latter systems dealt with inert, non-responsive crowders while in our homo-crowding case the system responds collective and self-consistent, i.e., the internal kinetics responds {\it in quorum} for all particles in the same way.
 
We finally present a perturbation theory, leading to scaling laws which describe the density-dependence of the switching kinetics quantitatively. For this, we consider the mean force on the particle size coordinate due to the interacting environment 
\begin{equation}
\label{eq.self.F_pp}
F(\sigma) = -\rho \int_V \mathrm{d}^3r \int_{-\infty}^{\infty} \mathrm{d}\sigma' N(\sigma') \frac{\partial \phi(r; \sigma, \sigma')}{\partial \sigma} g(r; \sigma, \sigma').
\end{equation}
This force is the source for the compression and tilt of the energy landscape $\psi$. In a first order perturbation approach to the spatial structure we apply the LDL $g(r; \sigma, \sigma') \approx \exp{\left[-\beta \phi(r; \sigma, \sigma')\right]}$. By integrating the force up to the particle's size, $\mathcal{F}(\sigma)=-\int_0^\sigma \mathrm{d}\sigma' F(\sigma') $, this results in a free energy of the analytical form linear in $\rho$ (with details in~\cite{SI})
\begin{eqnarray}
\mathcal{F}(\sigma) = \frac{5}{12}\pi\rho\epsilon\kappa \left(\sigma^3 + 3\sigma^2\left<\sigma\right> + 3\sigma\left<\sigma^2\right> \right),
\label{eqn:ApproxFreeEnergy}
\end{eqnarray}
where $\kappa \approx 6.377\cdot10^{-4}$ from numerical evaluation for $\beta\epsilon = 500$ in the Hertzian pair potential, and $\mathcal{F}(\sigma)$ can be conveniently expressed by the expectation values $\langle \sigma \rangle$ and $\langle \sigma^2 \rangle$ with respect to the emergent distribution. From another first order perturbation approach, $N(\sigma;\rho)\simeq p(\sigma)$ in Eq.~(\ref{eq.self.F_pp}), we obtain analytically $\langle \sigma \rangle = (\mu_1+\mu_2)/2$ and $\langle \sigma^2 \rangle = \delta^2 + (\mu_1^2 + \mu_2^2)/2$.  

Employing the modified free energy landscape $\tilde \psi(\sigma) \simeq \psi(\sigma)+\mathcal{F}(\sigma)$ in Kramers' theory, Eq.~(\ref{eqn:KramersRule}),  
%We assume that the position of the two minima ($\mu_{1,2}$) and the maximum ($\mu_3$) in the energy landscape do not shift in the $\sigma$-direction, which is justified for small perturbations, and validated by the simulations (Fig.~\ref{fgr:fig2}(b)). The activation energy can then be split into the part from intrinsic energy landscape $E_0^a$ (cf. Fig.~\ref{fgr:fig1}(a)), and a part resulting from the interactions. This leads to the
 we obtain following linear forms in $\rho$ for the activation energies, $E_\mathrm{SL}^a \approx \tilde \psi(\mu_3) - \tilde \psi(\mu_1) = E_0^a + \frac{\nu_\mathrm{SL}}{\beta}\rho\sigma_0^3$ and $E_\mathrm{LS}^a \approx \tilde \psi(\mu_3) - \tilde \psi(\mu_2) = E_0^a - \frac{\nu_\mathrm{LS}}{\beta}\rho\sigma_0^3$ with $\beta E_0^a = 1.019$, and $\nu_\mathrm{SL}=0.557$ and $\nu_\mathrm{LS}=0.697$ from evaluation of Eq.~(\ref{eqn:ApproxFreeEnergy})~\cite{SI}.  The theoretical results are shown through lines in Fig.~\ref{fgr:fig3}(b) and compare very well with the simulations. 

Applying Kramer's rule, $\tau^{\mathrm{FP}}\propto \exp(\beta E^a)$, now leads to MFPT predictions  
\begin{eqnarray}
 \tau^{\rm FP}_{\rm SL}(\rho)  \propto \left(\exp(\tilde\nu_{\rm SL})\right)^{\rho}  \;\;{\rm and}\;\; 
 \tau^{\rm FP}_{\rm LS}(\rho)  \propto \left(\exp(\tilde\nu_{\rm LS})\right)^{-\rho},
\end{eqnarray}
% \begin{eqnarray}
%  \tau_\mathrm{S\rightarrow L}(\alpha) &=& f(\alpha) \exp(\beta E_0^a) \exp(\nu_\mathrm{SL}\rho) \propto \left(\exp(\nu_\mathrm{SL})\right)^\rho, \\
%  \tau_\mathrm{L\rightarrow S}(\alpha) &=& f(\alpha) \exp(\beta E_0^a) \exp(-\nu_\mathrm{LS}\rho) \propto \left(\exp(\nu_\mathrm{LS})\right)^{-\rho}.
% \end{eqnarray}
with $\tilde \nu_{\rm SL} = \nu_{\rm SL}\sigma_0^3/\beta$ and $\tilde \nu_{\rm LS} = \nu_{\rm LS}\sigma_0^3/\beta$, establishing exponential scaling laws for the dependence of the MFPT on the packing density. As with the activation energies, the theoretical predictions for $\tau^{\rm FP}$ are in very good agreement with the computational results, cf. Fig.~\ref{fgr:fig3}(c).  From the nature of the perturbation theory, the scaling laws are quite universal for any pair potential in the LDL. In our case, however, it works up to surprisingly high densities. The reason lies on one hand in the softness of the Hertzian pair potential, featuring only weakly correlated RDFs, cf.~Fig.~\ref{fgr:fig2}.  On the other hand, our parent distribution is relatively broad, perturbing the energy landscape of the internal DOF only linearly in leading order, cf. Eq.~(\ref{eqn:ApproxFreeEnergy}).  

Our work provides a general methodical framework for studying the effects of (homeo)-crowding on the internal kinetics in complex responsive systems; this may include multi-component mixtures, complex (polymodal) energy landscapes, and various internal DOFs. It will be fascinating to add internal {\it activity}~\cite{Hartmut} into the RC model to study actively self-regulated crowding effects.  For example, internally fueled switching of hydrogel properties~\cite{Heuser, oscillating, DNA_hydrogel} are expected to lead to complex collective nonequilibrium effects~\cite{moncho,moncho2} and may be programmable to mimic the quorum sensing displayed by bacterial populations~\cite{quorum,Andreas}. 

\begin{acknowledgments}
The authors acknowledge support for computing facilities provided by the High Performance Cluster framework by the state of Baden-W{\"u}rttemberg (bwHPC) and the German Research Foundation (DFG) through grant no. INST 39/963-1 FUGG (bwForCluster NEMO).\end{acknowledgments}

%apsrev4-2.bst 2019-01-14 (MD) hand-edited version of apsrev4-1.bst
%Control: key (0)
%Control: author (8) initials jnrlst
%Control: editor formatted (1) identically to author
%Control: production of article title (0) allowed
%Control: page (0) single
%Control: year (1) truncated
%Control: production of eprint (0) enabled
%

%\bibliographystyle{apsrev4-2}
%\bibliography{bimodal,rpSim1}% Produces the bibliography via BibTeX.
 
\end{document}